# Obstacles to mathematization in introductory physics

*Suzanne White Brahmia, Rutgers University*
*Andrew Boudreaux, Western Washington University*
*Stephen E. Kanim, New Mexico State University*

**Abstract**

University students taking introductory physics are generally successful executing mathematical procedures in context, but often struggle with the use of mathematical concepts for sense making. Physics instructors note that their students experience difficulty with basic algebraic reasoning, a foundation on which more advanced mathematical thinking rests. However, little systematic research has been done to measure and categorize such difficulties in this population. This paper describes an investigation of trends in student reasoning with proportions, quantification, and symbolizing in introductory calculus-based physics. Although the assessment items used to probe student thinking require mathematical reasoning typically taught at the middle school level in mathematics courses, we find success rates of about 50% among calculus-based physics students. For many of these students, numerical complexity and the level of abstraction of the quantities interfere with basic arithmetic reasoning. We argue that the algebraic thinking of physicists stems from an idiosyncratic cognitive blend, which is not addressed in prerequisite algebra courses. We suggest that for more students to understand and adopt the mathematical thinking characteristic of physics, instructors and education researchers must explore how to make mathematization a more explicit part of the curriculum.

## I. Introduction

*Mathematizing* in physics involves translating between the physical world and the symbolic world in an effort to understand how things work.[1,2] Specific skills include representing concepts symbolically, defining problems quantitatively, and verifying that solutions make sense. Physicists develop and communicate ideas through the shared meanings they have built around these strong connections between mathematics and physics.

Arithmetic and algebraic reasoning are cornerstones of mathematization in an introductory physics course. Although students in a calculus-based course will have successfully completed prerequisite algebra courses, experienced instructors recognize that even their well-prepared students can struggle with algebraic decision making in physics. Whether it is the naïve but common association of negative acceleration with decreasing speed, or the sometimes mistaken inference that the $F_{net} = 0$ condition corresponds to an absence of forces, basic mathematization poses challenges to students throughout introductory physics.

Physics curricula typically rely on flexibility with algebraic reasoning that is special to physics and deeply embedded in the discipline.[3,4] While many investigations have focused on how physics student use mathematics at the level of pre-calculus and above, little research has examined the challenges students face thinking arithmetically in a calculus-based physics course. This paper extends work into this area by identifying specific reasoning difficulties in a large sample of mathematically well-prepared university physics students. We have administered written questions to investigate facility in the mathematical cognitive domains of generalized structural reasoning, quantification, and symbolizing. Our results suggest that standard instruction results in limited student capacity for basic mathematical sense making in physics. The following questions have guided our research.

After one semester of introductory calculus-based physics, to what extent are mathematically well-prepared engineering students likely to successfully reason with:

- ratio and proportion?
- the value and units that comprise *quantity*, across levels of numerical complexity and physical contexts?
- variable quantities in arithmetically simple situations?

The first question probes structural reasoning with ratio, a mathematical object common in the quantitative definitions of physical quantities (*e.g.*, velocity) and in physical laws *(e.g., a = F/m)*. We have examined how flexibly students apply ratios, and whether students think generatively with ratios to characterize novel situations. The second research question explores quantification, by assessing the extent to which students interpret and construct physically meaningful physical quantities in context. Finally, the third question investigates students' symbolizing, *i.e.*, using symbols to represent generalized numbers and relationships when reasoning quantitatively.

We approach both the design of assessment items and analysis of student responses from a cognitive blending framework,[5,6] treating the mathematics and physics as a single thinking space. The following section describes this framework. Section III describes how productive mathematical thinking in physics relies on generalized structural reasoning, quantification and symbolizing, summarizes relevant prior research in these areas, and presents the assessment items used in this study. Sections IV and V present our research methods and results; we find that generalized structural reasoning, quantification, and symbolizing present substantial obstacles to the development of introductory physics students' mathematization, and that surface features of problem context and numerical complexity can interfere with the reasoning of even well-prepared students. Section VI discusses implications for instruction.

## II. Theoretical Framework: Cognitive Blending

The theoretical framework of cognitive blending [5,6] is consistent with our view that continuous interdependence of thinking about the mathematical and physical worlds is necessary for expert problem solving in physics. Figure 1 illustrates a *double scope arithmetic reasoning blend*, in which two distinct domains of thinking are merged to form a new cognitive space that is optimally suited for productive work.

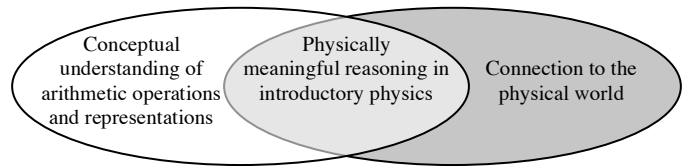

**Figure 1:** Double scope arithmetic reasoning blend

Prior work in this area presents a theoretical framework that spans a spectrum of homogeneity of the cognitive blend (represented as the overlap in Fig. 1). Researchers who focus on students' use of relatively sophisticated mathematics [3,4,7] commonly model the mathematical world and physical world as a more heterogeneous thinking space, as students grapple with multistep mathematical operations within physics contexts. Considering the use of algebra at the introductory level, Tuminaro also describes heterogeneity: "students invoke ideas from mathematics—such as equations, graphs, etc.— to help them understand the physics."[8] In contrast, the case study research of Hull *et al*. hypothesizes a more homogeneous blend of these thinking spaces in the context of introductory physics.[9] From the mathematics research perspective, Czocher conducted a microscopic study with engineering students enrolled in a differential equations course and observed them solving a variety of physics problems over the course of the semester.[10] She reports that successful students functioned most of the time in a "mathematically structured real-world" in which they moved back and forth fluidly between physics ideas and mathematical concepts. Czocher describes this thinking space as being between the "real world" and the "math world."

We suggest that student learning in introductory physics is best supported through a completely homogeneous blend (as observed by Czocher), such that there is no distinction between the physics and the arithmetic worlds. We propose a thinking space we refer to as the *mathematization of introductory physics*, in which physical sensemaking is essential for and integrated with mathematical reasoning. In the context of arithmetic thinking we claim that the optimum thinking space is a heterogeneous blend representing a continuous interdependence between the physical world and conceptual understanding of arithmetic operations and representations. We analyze our results using this framework and draw conclusions that can inform both instruction and curriculum development.

### III. Underpinnings and Assessment Items

A growing body of literature from mathematics and physics education research documents student and expert use of mathematics in physics at the level of pre-calculus and above.[7, 10-19] Caballero, Wilcox, Doughty and Pollock characterize research involving upper-division physics students using two broad categories: 1) *macroscopic*, or whole class studies focused on uncovering student difficulties, and 2) *microscopic*, typically theory-driven studies, focused on in-situ interviews with small groups of students.[20] We agree with Caballero *et al.* that a more complete understanding emerges from connecting these approaches.

We extend the distinction articulated in Caballero *et al.* to include research on mathematical sensemaking in introductory physics. Tuminaro, in his microscopic study of students in algebra-based introductory physics courses, points out that if students do not expect conceptual knowledge of mathematics to connect to their work in solving physics problems, then they are likely to frame their problem-solving activities in terms of plug-and-chug manipulations or intuitive sense-making that is primarily qualitative.[8] He concludes that for these students, sense making is not part of calculating. Hull, Kuo, Gupta and Elby have seen similar results in their microscopic study.[9] Our macroscopic study of student difficulties compliments these microscopic studies, contributing to a more complete understanding.

Fewer than 3% of the students who take the introductory course go on to major in physics, so most introductory students are not represented in studies of upper-division students. Estimating the prevalence of introductory students' difficulties with mathematical sensemaking will increase overall understanding of student mathematization in physics. This paper reports on a carefully constructed, large-N study of introductory, calculus-based physics students to document reasoning trends that characterize this population. This work can be considered a macroscopic study with theoretical underpinnings, intended to uncover the extent to which students struggle with some of the algebraic ideas foundational to introductory physics.

Just as with upper division courses, sensemaking at the introductory level involves creative mathematical thinking. In contrast to procedural use of math, generating algebraic descriptions of physical events and systems requires students to try approaches without knowing whether or not they will work, which in turn requires courage and tolerance of failure. Students must learn to check whether or not the mathematics they generate makes sense, and how to iterate toward better solutions. We view the three mathematical cognitive domains of structural reasoning, quantification and symbolizing as building blocks for productive mathematical thinking, and as a foundation for the sophisticated algebraic and calculus reasoning ubiquitous beyond first year physics. Below we describe each of these areas by summarizing prior work and illustrating how difficulties might impact physics learning. We also present assessment tasks we have developed to probe student thinking.

A. Generalized structural reasoning

In order to connect mathematics to physical phenomena, physicists often make use of the properties of common mathematical structures. A generalized mathematical structure (*e.g.* a $1/r^2$ force), when recognized, can guide thinking in a new context. For example, when studying orbits in intermediate mechanics, a student might recognize from their understanding of the Coulomb interaction that the attractive $1/r^2$ force implies that the potential energy is finite even at infinite distance, and that therefore it should be possible for a satellite to "escape" a center of gravitational attraction.

While common practice for physicists, this way of thinking is novel for introductory physics students. In his microscopic study of engineering students at a highly selective university, Sherin reports on the opacity of the mathematical structures underlying the kinematics equations.[17] He posed the following task to students in a third-semester introductory calculus-based physics course: *"Imagine that we've got a pile of sand and that, each second, R grams of sand are added to the pile. Initially, the pile has P grams in it. Write an expression for the mass of the sand after t seconds."* In clinical interviews, all students were able to generate a correct expression for the mass as a function of time. None, however, recognized that the arithmetic progression for the sand pile mass matched that of the velocity function *v(t)* for a motion with constant acceleration. The interview subjects could not explain why the "correction" to $v_o$ should be *at*, and seemed perplexed to be asked to consider such a simple question about sand.

Rebello, Cui, Bennett, and Zollman report on students' capacity to generalize the reasoning and methods learned in trigonometry and calculus to end-of-chapter textbook problems in physics.[16] Calculus-based physics students were asked to solve physics problems involving simple integration or differentiation similar to problems they had already solved in homework. While the students were able to execute the required calculus procedures when prompted, they were largely unsuccessful at setting up and solving problems that required them to select appropriate calculus tools and adapt them to fit a physical situation. The researchers also surveyed algebra-based students before and after instruction and found little evidence that students would spontaneously generalize trigonometry from math to physics; students lacked flexibility with the prerequisite mathematics. Schoenfeld describes math students' *belief systems* (*i.e.* expectations) as rigid due to their emphasis on the context-specific nature of problem-solving approaches (*e.g.*, they use deductive argumentation in geometry proofs but not in other contexts).[21]

We associate the observed lack of spontaneous generalizing to rigidity in students' beliefs about how and when mathematics should be used. In particular, student beliefs may not allow for spontaneous uses of mathematical reasoning unprompted outside of math class. We find it productive to think in terms of Hammer's resource framework[22,23] and Wittmann's coordinated set of resources.[24] Rebello's students do not activate calculus or trigonometry resources in the physics context unless they are explicitly prompted to do so, nor do Sherin's, even when they are prompted via an analogy. Each example above describes a reasoning structure that experts readily activate in a variety of contexts. We think of the reasoning structure as composed of a coordinated set of resources, which in turn requires a robust set of individual resources to be readily accessible. In the Sherin study, the activation of students' coordinated arithmetic progression set was perhaps context dependent. While sand was a context that activated the reasoning, without scaffolding the coordinated set was not broad enough to include physics quantities, and students did not generalize the reasoning.

We view ratio as one of the most important general mathematical structures in introductory physics. Student reasoning about ratio and proportion was examined in the early physics education work of Arons, Karplus, and others,[25-28] as well as extensively in mathematics education research.[29-32] Thompson describes proportional reasoning as interconnected skills that are context-dependent, claiming that proportional reasoning "appears in various guises in different contexts and different levels of sophistication."[33] The assessment items reported on in this paper stem from our work on a related, ongoing project, in which we are engaged in delineating and assessing some of the specific skills that make up proportional reasoning in physics.[34]

Table I presents some of the items used in the study described in this paper to investigate our first research question, associated with generalized structural reasoning. These items require students to either apply a given ratio (items I and II), or identify ratio as an appropriate measure (items III and IV). Item III is modified from a microscopic study of pre-service elementary teachers' understanding of ratio-as-measure,[35] and Item IV is a physics-y version. Students who have internalized ratio as a general mathematical structure will have a powerful resource for determining how to use the relevant ratio appropriately, and for checking the validity of their answer. In contrast, poor performance would suggest lack of an internalized ratio structure.

**Table I:** Items used to assess generalized structural reasoning with ratio.

| Item name | Item text |
|---|---|
| I. Olive Oil | You go to the farmer's market to buy olive oil. When you arrive you realize that you have only one dollar in your pocket. The clerk sells you 0.26 pints of olive oil for one dollar. You plan next week to buy 3 pints of olive oil. Which of the following expressions helps figure out how much this will cost (in dollars)?<br>a. 3/0.26   b. 0.26/3   c. 3•0.26   d. (3+1)•0.26   e. none of these |
| II. Traxolene | You are part of a team that has invented a new, high-tech material called "traxolene." One gram of traxolene has a volume of 0.41 cm$^3$. For a laboratory experiment, you are working with a piece of traxolene that has a volume of 3 cm$^3$. Which of the following expressions helps figure out the mass of this piece of traxolene (in grams)?<br>a. 3/0.41   b. 0.41/3   c. 3•0.41   d. (3+1)•0.41   e. none of these |
| III. Square Buildings | You are riding in an airplane. Below you see three rectangular buildings with the rooftop dimensions shown at right. You are interested in how close the shapes of the rooftops of the buildings are to being square. You decide to rank them by "squareness," from *most* square to *least* square. Which of the following choices is the best ranking?<br>a. A, B, C   b. B, A, C   c. C, A, B   d. C, B, A   e. B, C, A<br><br>*Building A: 77 ft by 93 ft*<br>*Building B: 51 ft by 64 ft*<br>*Building C: 96 ft by 150 ft* |
| IV. Force Vectors | Each of three different objects (A, B, C) experience two forces, one in the +x direction and one in the +y direction. Rank each object according to how close the direction of the net force is to a 45° angle between the x-direction and the y-direction, from *closest* to 45° to *farthest* from 45°.<br>a. A, B, C   b. B, A, C   c. C, A, B   d. C, B, A   e. B, C, A<br><br>\| \| Force in x-direct \| Force in y-direct \|<br>\| A \| 77 N \| 93 N \|<br>\| B \| 51 N \| 64 N \|<br>\| C \| 96 N \| 150 N \| |

While each item I-IV stands on its own as an assessment of structural reasoning about ratio, the items together form pairs of questions that can be compared to probe the context dependence of student reasoning. Items I and II are isomorphic, and involve identical arithmetic reasoning. Only the surface features vary, with item I involving an everyday context (purchasing olive oil at the market) and item II, a physics context (a high-tech material called traxolene). Each question can be answered, however, by using a characteristic ratio to find an unknown amount. Items III and IV are isomorphic in a similar fashion.

### B. Quantification

Quantification is fundamental to reasoning in physics, and physical quantities are the objects of this reasoning. Take for example a ball rolling across the floor, quantification involves thinking of measurable quantities that can help describe the motion (e.g. rates, distances), their units and the arithmetic involved in constructing them. Researchers in mathematics education have identified *quantification* as a significant challenge to students who are learning to mathematize. Thompson, who has researched and written extensively on this topic over the past two decades, defines quantification to be " the process of conceptualizing a mathematical object and an attribute of it so that the attribute has a unit of measure, and the attribute's measure entails a proportional relationship … with its unit." He considers quantification to be "a root of mathematical thinking", and argues that learners develop their mathematics from reasoning about quantities. [33,36]

In a typical introductory university course, students encounter $\sim 10^2$ new physical quantities. A physical quantity involves a value and an associated unit (and sometimes a direction in space). Physicists commonly use all features of quantity to guide their own thinking about relationships between quantities, and even to formulate new ones. In order to begin to quantify efficiently in physics, students must have a conceptual facility with value of all kinds, including large and small numbers and general variables, as well as a facility reasoning with units. As more abstract units are introduced, the interpretation becomes more challenging. (Consider, for example, a compound unit such as $m/s^2$, which combines length and time in a complicated way. The meaning of "$s^2$" is not readily evident.) Facility with value and unit supports the development of conceptual understanding of the arithmetic involved in combining quantities. Introductory physics introduces a new challenge with vector quantities and the specific algebra they obey.

Mastery of numeric value may seem trivial and well outside the domain of college physics. But *mastering number in the context of physical quantities* is not the same as mastering number in math class, where units are rarely involved. In physics, units carry deep meanings. These meanings often contrast with common uses of numbers in everyday life, which include categorizing ("What's behind door number 1?"), ordering ("My amp goes up to 11!") and defining thresholds (a blood pressure of 120/80 means don't worry). Physicists work fluidly at the interval level of measurement, where the quantity itself (the numeric value and its associated unit) carries important information. For example, a steady speed of 25 mph immediately conveys that the vehicle will travel 25 miles in one hour, while students may think about 25 mph simply as "slower than I want to drive."

Prior research has shown that the complexity of numeric values produces cognitive strain that can affect basic arithmetic decision-making.[37,38] We hypothesize that this may also be true for students as they learn physics. Measured quantities more commonly involve decimals and fractions than whole numbers, whereas whole numbers are more common in school algebra. Numbers very large or small compared to everyday values, which are ubiquitous in physics, also pose difficulty and are not frequently used in algebra courses. We explore the effect of numerical complexity on arithmetic decision making in this study.

Physics also involves compound quantities that result from multiplying and dividing other quantities (e.g., momentum). While the arithmetic procedures involved in creating new quantities are not necessarily challenging, deciding when and why this arithmetic makes sense can be difficult for students.[36] Interpreting and understanding ratios or products involve conceptualizing multiplication and division. Many compound quantities are rates of change, which require a conceptual understanding not only of ratios but also differences. Conservation principles, which require an understanding of "net amount," are a common motivation for the development of a new quantity. Some quantities, such as electric flux, combine quantities already poorly understood (electric field, area). Tuminaro has reported on student difficulties conceptualizing arithmetic and the simplest multiplicative structures in physics. [8]

Unlike working with pure numbers in math class, special constraints apply when adding, subtracting, multiplying, and dividing quantities in physics. For example, addition and subtraction can be carried out only with like quantities expressed in like units, while multiplying and dividing can create a quantity completely different from either of the constituents. Students have little or no experience reasoning about multiplicative structures with physics quantities from prerequisite math classes, but are expected to reason this way in physics. For example, a student might be expected to recognize from context whether the product of a force and a distance yields a torque or work. The research described in this article explores how physics contexts might pose challenges to student's arithmetic reasoning.

Table II presents items we have used to investigate our second research question, involving student quantification. Items V and VI require students to interpret an unfamiliar ratio by attaching a specific meaning to the units that result from dividing one physical quantity by another. On items VII a and b, students must create a quantity by constructing a ratio to match a given physical interpretation. Students who recognize that a physical quantity is instantiated by a numerical value linked to an associated unit will be more likely to succeed on these tasks.

**Table II:** Items used to assess quantification.

| Item name | Item text |
|---|---|
| V. Paint | Catherine is hired to paint the ceiling of her aunt's living room. She covers the ceiling with a uniform coat of paint. The ceiling has a surface area of 580 square feet. After finishing, Catherine notes that she used 2.4 gallons of paint. Catherine divides 580 by 2.4 and gets 241.7.<br>Which of the following statements about the number 241.7 is true?<br>a. 241.7 is the total number of gallons of paint used<br>b. 241.7 is the total number of square feet of surface area covered by the paint<br>c. 241.7 is the number of gallons of paint that covers one square foot<br>d. 241.7 is the number of square feet that one gallon of paint covers<br>e. none of the above |
| VI. Door Knob | Catherine shuffles her feet across her living room carpet and then she touches a doorknob, which has a surface area of 580 square centimeters. When she touches the doorknob she transfers 2.4 microcoulombs of electric charge that spreads out uniformly over the doorknob's surface. Catherine divides 580 by 2.4 and gets 241.7.<br>Which of the following statements about the number 241.7 is true?<br>a. 241.7 is the total number of microcoulombs of charge transferred<br>b. 241.7 is the total number of square centimeters of surface area covered by the charge<br>c. 241.7 is the number of microcoulombs of charge that covers one square centimeter<br>d. 241.7 is the number of square centimeters that one microcoulomb of charge covers<br>e. none of the above |
| VIIa. Rice – Whole | Bartholomew is making rice pudding using his grandmother's recipe. For three servings of pudding the ingredients include 4 pints of milk and 2 cups of rice. Bartholomew looks in his refrigerator and sees he has one pint of milk. Given that he wants to use all of the milk, which of the following expressions will help Bartholomew figure out how many cups of rice he should use?<br>a. 4/2   b. 2/4   c. 2•4   d. (2+1)•4   e. none of these |
| VIIb. Rice – Decimal | Same as VIIa except decimal quantities are used<br>"… include 0.75 pints of milk and 0.5 cups of rice…"<br>a. 0.5/0.75   b. 0.75/0.5   c. 0.5•0.75   d. (0.5+1)•0.75   e. none of these |

Items V and VI constitute a matched pair, identical in the underlying mathematical reasoning, but different in surface features, allowing the context dependence of student reasoning to be probed. Items VII a and b are identical save for differences in the way the quantities are represented, with version a) involving whole numbers, and version b), decimal numbers. This allows the impact of numerical complexity on student quantification to be examined.

C. Symbolizing

Mathematics education researchers have been investigating symbol use for decades.[39] The context dependence of symbol use in physics is nuanced, and often not part of students' mathematics preparation. Below we explore roles that fundamental symbols – variables and the equals sign– play in the development of physics concepts. Assessment items and empirical findings reported later in the article focus on these symbols.

Extensive research on student use of variables has identified persistent difficulties.[40-42] These difficulties can be compounded in physics contexts: physicists use symbols in idiosyncratic ways that may confuse students. At the beginning of many mathematics textbooks there are lists of letters that are to be considered variables *(x, y, z)* and of letters to be considered constants *(a, b, c)*. Typically, an expression will contain only a single variable; the task at hand is usually to solve for that variable.[43] In contrast, physicists are more fluid in their use of symbols, and the same letter might be a constant in one problem, and a variable in another. Even a physical constant can be treated as a variable under certain conditions. For example, students may be asked to "find the value of *g* on this planet," or even to "take the derivative with respect to *h-bar*."

In addition to difficulties with the variables that make up an equation, students may struggle with the nuances of the equals sign involved in physics. High school and college students may use the equals sign inappropriately as they solve equations or evaluate expressions in algebra and calculus.[44] While many students can interpret an equals sign as a prompt for calculation (e.g., $v$ = 3 m/s × 5 s), fewer understand it to be the relational symbol of mathematical equivalence (e.g., $(\boldsymbol{F}_{a \text{ on } b} + \boldsymbol{F}_{c \text{ on } b})/m_b = \boldsymbol{a}_b$).

Cohen and Kanim, building on early work by Clement, Lochhead and Monk used the "students-and-professors" question to probe student ability to convert a natural language sentence into a mathematical expression.[45,46] Students were asked to write an equation, using *S* for the number of students and *P* for the number of professors, to represent the statement, "There are six times as many students as professors at this university." Clement *et al*. report that students taking calculus-based introductory physics found this task challenging, commonly placing the number 6 on the wrong side of the equation. Cohen and Kanim explored this "reversal error" in greater detail by changing sentence structure and the choice of symbols, and found that about two-fifths of students making the error seemed to be performing a word-order translation of the sentence (referred to as *syntactic translation)*, while most of the remaining students seemed to be treating the symbols *S* and *P* as units or labels, rather than variables (*i.e.,* interpreting the expression "6*S*" to mean "there are six students").

The equations typically encountered in physics courses are far more symbol rich than the equation involved in the students-and-professors question. In a macroscopic study at the University of Illinois, Torigoe and Gladding posed isomorphic questions on final exams in the introductory physics course for engineers.[47] One question in the pair used only numbers, while its partner used only symbolic values. Differences in success rates of up to 65% were observed. Student success on exam questions that relied on accurate manipulation of uniquely symbolic representations correlated to course grades, with the strongest correlation occurring for the bottom quartile of the students. Students who reason poorly when faced with a host of symbolic quantities seem likely to struggle in an introductory physics course.

Torigoe and Gladding's results indicate that use of purely symbolic values in the statement of a physics problem can impact student performance. Our research explores whether representing just a single quantity with a variable can impede students' reasoning. Table III presents items we have used to investigate symbolizing flexibility.

**Table III:** Items used to assess symbolizing.

| Item name | Item text |
|---|---|
| VIIc. Rice – Fraction | Same as Table 2, VIIa except a fractional and a decimal quantity used <br> "… include 0.75 pints of milk and 5/8 cups of rice…" <br> a. (5/8) /0.75   b. 0.75/(5/8)   c. (5/8) • 0.75   d. ((5/8) +1) • 0.75   e. none of these |
| VIId. Rice – Variable | Same as Table 2, VIIa except a fractional and a variable quantity is used <br> "… include $N$ pints of milk and 5/8 cups of rice…" <br> a. (5/8)/$N$   b. $N$/(5/8)]   c. $N$ x (5/8)   d. ((5/8)+1) x $N$   e. none of these |
| VIII. Woozles | Consider the following statement about Winnie the Pooh's dream:  *"There are three times as many heffalumps as woozles."*  Some students were asked to write an equation to represent this statement, using $h$ for the number of heffalumps and $w$ for the number of woozles.  Which of the following is correct? <br> a. *3h/w*    b. *3h = w*    c. *3h + w*    d. *h = 3w*    e. both a and b |

Item VIII requires students to represent a proportional relationship with an algebraic statement.  Items VII c and d require analogous reasoning to that required for items VII b and a, but with one of the quantities represented by a variable.

## IV. Research methods

In order to uncover specific challenges that students encounter, we administered the multiple-choice questions described in the previous section at the beginning and end of introductory physics and chemistry courses for engineering students taught at Rutgers University, a large public research university. In this section, we describe the student population and methods of data collection and analysis.

### A. Development of research tasks

The research tasks were drawn from a large set of items used by the authors in an investigation of the proportional reasoning of introductory physics students.  The development and validation of the items is described in detail elsewhere;[34]  here we summarize the process. Rather than procedural or computational skill, the questions focus on sensemaking and conceptualization of ratio quantities. The initial versions of the items asked students to explain their reasoning and show their work. Question validity was established through in-depth, think-aloud interviews with more than twenty individual students. We used the written responses and interview transcripts to create multiple-choice versions of the questions, with distractors based on the difficulties identified through analysis of students' verbal explanations. Because the current macroscopic study identifies trends in large populations of students, we focus on quantitative results from the multiple-choice versions.

Throughout the development of the items, we observed variations in student reasoning associated with physical context, with the level of abstraction of the ratio or product quantity, and with the numerical complexity of the quantities involved.  These findings, consistent with previous studies of student reasoning about ratio,[27, 28, 48] led us to develop parallel versions of several of the assessment questions. We hoped to isolate triggers for variations in student reasoning by changing only a single surface feature of each question.

For the current study we selected items, described and shown in the previous section in tables I-III, that involve the mathematical cognitive domains described earlier:  generalized structural 1010reasoning, quantification, and symbolizing. Due to the interrelated nature of these cognitive domains, it is not possible to design items that target one domain at the exclusion of the others. We thus present items that *highlight* the cognitive domain of interest, acknowledging that the other domains may also be relevant to student responses.

B. Student population

This study's population is freshman non-honors engineering students at Rutgers University taking traditionally taught calculus-based physics and chemistry in the same semester. These students tend to be well prepared mathematically, with a mean mathematics SAT (2011/2012 test version) score of 680. The data was collected as part of routine course pre and post testing, which includes concept inventories in addition to a suite of questions associated with ratio reasoning. Throughout the semester, the lecturer in the physics course modeled proportional reasoning (and other mathematical methods) in the context of the physics content being taught.

C. Data collection

We administered the research tasks under exam conditions as an ungraded in-class pretest during the first week of the introductory, calculus-based physics course and the general chemistry course in Fall 2013. In the physics course, the same tasks were administered again as a posttest, ten days before the end of the semester, also under exam conditions. In the chemistry course, however, the post-test was administered online outside of class, and there was a substantial drop in the number of students participating. Table IV summarizes the administration of pre- and post-test questions.

**Table IV.** Administration of written assessment items reported on in this paper.

| Class | Subject | Items reported on in this paper | Pretest | Posttest | Versions of test |
|---|---|---|---|---|---|
| Freshmen | Intro physics | III, VII a-d, VIII | Supervised In class | Supervised In class | 8 |
| Freshman | Gen Chem | I, II, IV, V VI | Supervised In class | Unsupervised Online | 6 |

In all, 14 multiple-choice items probing different facets of proportional reasoning were administered on the pretest and again on the posttest. Seven items were administered in the mechanics course ($n_{pre}$=770 and $n_{post}$=737), and seven in the chemistry course ($n_{pre}$=628 and $n_{post}$=332). A subset of the students took both the mechanics and the chemistry tests (479 on the pretests and 287 on the posttests). In both courses, the items were bundled with a standardized concept inventory (the Force Concept Inventory [49] in the physics course and the Chemical Concepts Inventory [50] in the chemistry course). In a single sitting, students first completed the proportional reasoning items, and then immediately completed the concept inventory. The students were not constrained by time and were awarded credit for participation. Note that these considerations apply to both the pretest and posttest, which were administered under identical conditions (except that in chemistry, the posttest was given online, see Table IV).

As mentioned in section IIIA, we probed the effect of surface features on student reasoning by administering matched versions of items on different versions of the tests. Tests were administered in the recitation section of the course, and within a given recitation different test versions were assigned randomly. Thus, for a given isomorphic question pair, half of the students in a given course received one version of the question and half received the other. Each student in the study received the same version of the question suite on the pretest and the posttest.

D. Data analysis

In this study, data collection and analysis has involved three types of comparisons:

- a single question administered to each student in the sample at two different times (*i.e.*, pre- and post-instruction),
- paired questions administered simultaneously to each student in the sample, and
- paired questions administered simultaneously to two different samples taken from the same population.

To measure performance in the first two comparisons, we use the McNemar test of significance, while in the third, we use the Mann-Whitney test. For the latter, we established baseline equivalence between samples using FCI pretest and SAT Math scores, with effect sizes associated with differences in the standard error found to be less than 0.5 on each.

## V. Results and discussion

Below we present and discuss responses to the multiple-choice items sequentially to evaluate student facility with generalized structural reasoning, quantification, and symbolizing. The section concludes with a summative discussion involving the cognitive blending framework.

A. Generalized Structural Reasoning

Items I-IV (shown in Table I) assess student ability to apply structural reasoning about ratio quantities. Results are summarized in Table V.

**Table V.** Results on items that assess generalized structural reasoning.

| Item: | I. Olive Oil[1] | II. Traxo-lene[1] | III. Square Buildings[2] | IV. Force Vectors[1] |
|---|---|---|---|---|
| N: | 155 | 177 | 275 | 275 |
| Correct: | 55% | 66% | 17% | 23% |
| **Comparison:** | I to II | | III to IV | |
| p-value: | .06 | | .05 | |
| Effect size: | N/A | | 1.9 | |

1: Administered as online posttest in chemistry course.
2: Administered as inclass posttest in physics course.

The mathematical reasoning for items I and II involves a single step that can easily be checked using dimensional analysis, yet over one-third of the engineering freshman answered incorrectly. The most common incorrect response on item I (0.26•3, choice c), given by nearly one-quarter of the students, corresponds to multiplication of a number of pints of olive oil and the olive oil cost in pints per dollar. Even without proportional reasoning, a student could recognize that such multiplication yields a quantity measured in the physically meaningless units of square pints per dollar. It seems that many students failed to apply ratio reasoning or even to use dimensional analysis for sense making. A similar interpretation applies to results on the Traxolene question (item II).

Items III and IV each involve creating a ratio as the appropriate measure for making a judgment about a specific physical situation. On item III, which involves judging the "squareness" of rectangular buildings, the difference in lengths of the two sides of the rectangle is meaningful only in comparison to the absolute lengths of those sides. A dimensionless ratio of the lengths thus serves as the appropriate comparison, and is independent of both the units used and the overall size of the rectangle. Only 17% of the freshman engineering students selected the ratio-based answer, however, while more than two-thirds selected an incorrect comparison based on the difference between the sides. For item IV, the Force vectors question, a student can recognize that the closer the ratio of the x- and y-components of a vector is to unity, the smaller the deviation of the direction of the vector is from 45°. Less than one-quarter of the students answered correctly. One-half of the students gave the difference-based response.

The lower rows of Table V compare responses on the two pairs of isomorphic questions (I & II, and III & IV). In each pair, the two items involve reasoning that an expert recognizes as identical. The first item uses an everyday context while the second uses a "physics" context. Student performance was slightly stronger on the physics context item in comparison to the item using an everyday context (e.g., students were more successful applying a ratio in the olive oil context than in the traxolene context). The differences, however, were only marginally significant, with p-values ~ 0.05.

Analysis of the free-response versions of the items indicates that many students believed they were being asked to recall a formula. In interviews, such students spent substantial time trying to remember such formulas; similar to the findings of Von Korff et al.,[19] many students seemed to privilege the authority of formulas over the reasoning needed for sense making. Students did not typically make use of the units associated with physical quantities to guide their thinking.

The stronger performance on the physics context versions of the isomorphic item pairs was unexpected. We predicted that everyday contexts would cue reasoning resources not triggered by a physics context, and speculate that an "equation authority" effect may be involved. In the first pair, item II involves the well-memorized formula for density ($d = m/v$). In interviews, students readily recalled this formula and used it. In contrast, a formula may not readily come to mind for the olive oil context of item I. While rote use of formulas can be problematic, formula use triggered in the physics context but not the everyday context possibly contributed to stronger performance. With readily accessible and strongly linked resources for structural reasoning with ratios, we might expect that even in the absence of a formula on the Olive oil question, students would recognize the proportional relationship and be guided toward a correct response. These results thus suggest that students' structural reasoning with ratio is not robust.

On the second pair of items, students commonly performed extensive trigonometry calculations on item IV (Force vectors), but not on item III (Square buildings). (We emphasize, however, that students strongly favored a difference strategy over a ratio strategy on each item.) Although the students in our study had a stronger background in math and science, our results are consistent with the findings of Simon and Blume on the difficulties encountered by pre-service elementary teachers on the Square buildings question.[35] As in the case of the first pair (items I and II), increased use of mathematical formalism, even if disconnected from sense making, may have contributed to the higher correct response rate.

In aggregate, these results indicate that applying a given ratio, as well as spontaneously choosing to construct a ratio as a basis for comparison, are challenging modes of thinking for engineering students. These in-context, generative uses of mathematics, while central to physics, may not be reliable cognitive tools for many students. If structural reasoning with ratio poses a challenge, we can further speculate that reasoning about more complex mathematical structures (e.g., an inverse square law) may involve even greater levels of difficulty.

B. Quantification

Physics experts conceptualize *quantity* as a numerical value tightly linked with an associated unit. To investigate student quantification, we have posed questions in which students must either verbally interpret a given ratio (items V and VI), or construct an appropriate ratio from measured values (items VIIa and b). Table VI summarizes student performance on these items.

Item VI involves interpreting the ratio of the surface area of a doorknob to the electric charge distributed on it. (Although the students completed the question as part of a mechanics course, electric charge was a topic in their chemistry curriculum, taken concurrently with mechanics.) We expected students to interpret the resulting quantity as the number of square centimeters of area required for each microcoulomb of charge. This is a non-standard quantity, the inverse of the more common surface charge density ratio, making it difficult for students to answer correctly using a memorized definition. Less than 60% of students answered correctly.

**Table VI.** Results on isomorphic pairs of items that assess quantification.

| Item: | V. Paint[1] | VI. Door knob[1] | VIIa. Rice-Whole[2] | VIIb. Rice-Decimal[2] |
|---|---|---|---|---|
| N | 280 | 291 | 171 | 177 |
| Correct | 88% | 59% | Pre: 78% <br> Post: 75% | Pre: 60% <br> Post: 66% |
| Comparison | V to VI | | VIIa[3] to VIIb[3] | |
| p-value | < 10⁻⁴ | | Pre: < 10⁻⁴ <br> Post: .02 | |
| Effect size | 11.6 | | Pre: 5.3 <br> Post: N/A | |

1: Administered as online posttest in chemistry course.
2: Administered in-class in physics course.
3: p-value > .20 for pre-to-post comparison of single item.

Item V requires the same reasoning as item VI, and even involves identical numerical values. The context for item V, however, is a more familiar situation: an amount of paint applied to a wall. As shown in Table VI, performance on item V was significantly stronger than on item VI.

Items VIIa and VIIb are nearly identical. Both involve a recipe context in which students must construct a ratio to find the number of cups of rice for each pint of milk, given the total numbers of cups of rice and pints of milk. The items differ only in numerical complexity: version a involves whole number quantities (2 cups of rice and 4 pints of milk) while version b involves decimal quantities (0.5 cups of rice and 0.75 pints of milk). The correct response rate on item VIIa was significantly higher than that on VIIb; it seems that for some of the engineering students, reasoning arithmetically with decimal numbers presents an obstacle not present when reasoning with whole numbers.

Results from these questions demonstrate the extent to which reasoning about quantity is sensitive to surface features of the quantities involved. More abstract quantities (such as electric charge) seem to inhibit the reasoning students are successful with in less abstract contexts (such as paint covering a wall). While a physicist would likely regard the paint and doorknob questions as similar, Fig. 2 shows that performance on the doorknob question was substantially weaker, suggesting that many students have difficulty generalizing the relevant reasoning with quantity across contexts with different surface features. The results on this isomorphic pair suggests that while most freshman engineering students may, in a sense, "possess" the reasoning resources needed to interpret a ratio quantity in context, robust quantification is lacking in part because many students do not reliably include units in their reasoning.

Similarly, on Items VII a-c, performance varies with the complexity of the numeric values. Although the necessary reasoning is identical, Fig. 3 shows that performance on the decimal and fraction versions of the Rice question was significantly weaker than that on the whole numbers version. Students apparently are distracted from or are less likely to cue the appropriate ratio reasoning when presented with decimal and fractional values.

Our results reveal that it is not always the case that physics contexts are more challenging for students than everyday contexts. We observe that many contexts can be difficult, especially when they involve unfamiliar or abstract quantities – a result similar to what has been observed on the FCI by researchers using modified question contexts.[51,52] While the issues discussed in our study may be generalizable to contexts outside of physics, we consider them to be specifically relevant in introductory physics because of the important role that algebraic reasoning plays in its discourse, and the immediate and constant introduction of new and abstract quantities.

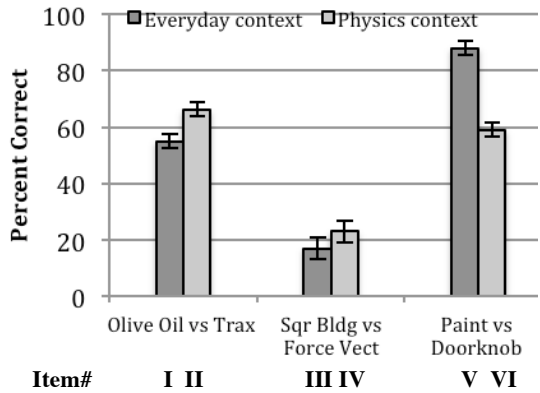

**Figure 2:** Context comparisons, pooled standard error shown for each comparison

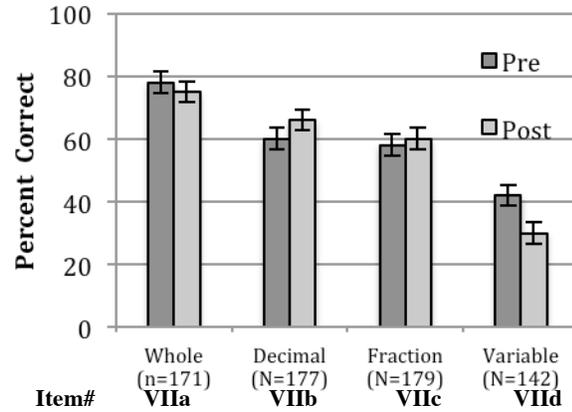

**Figure 3:** Rice questions, pooled standard error =3.4%

Quantification is cognitively challenging; unfamiliar units and complexities in the representations of value can derail students as they struggle to reason about the many new quantities they encounter in a physics course. Physicists use units and dimension to guide their reasoning about new physical quantities. This generative use of mathematics may be foreign to introductory physics students, and any nascent abilities may be overwhelmed in situations in which the complexity of the numbers or the level of abstraction of the quantities is high.

C. Symbolizing

Prior research has shown that students struggle with purely multi-variable expressions;[47] here we investigate whether student reasoning is affected by including only a single variable. Item VIIc represents relevant quantities with numbers (5/8 cups of rice and 0.75 pints of milk), while version d replaces the decimal quantity with a general variable "$N$" (5/8 cups of rice and $N$ pints of milk). Item VIII, taken from Cohen and Kanim,[45] probes symbolizing of a verbal statement that includes an equals sign. Table VII summarizes results on items VIIc, VIId, and VIII.

**Table VII.** Results on items that assess student ability to work with symbols in context.

| Item: | VIIc. Rice-Fraction[2] | | VIId. Rice-Variable[2] | | VIII. Woozles[2] | |
|---|---|---|---|---|---|---|
| N: | 179 | | 142 | | 685 | |
| Correct: | Pre: | 58% | Pre: | 42% | Pre: | 48% |
| | Post: | 60% | Post: | 30% | Post: | 49% |
| Comparison: | VIIc[3] to VIId[4] | | | | Pre/Post | |
| p-value: | Pre: | < $10^{-3}$ | | | 0.85 | |
| | Post: | < $10^{-9}$ | | | | |
| Effect size: | Pre: | 4.7 | | | N/A | |
| | Post: | 8.8 | | | | |

Notes:
2: administered in-class as part of physics course
3: p-value>.20 for pre-to-post comparison of single item
4: p-value=.04 for pre-to-post comparison of VIId with an effect size of 3.0

We see a significant and large difference on items VIIc and VIId, providing strong evidence that the presence of just one symbolic variable quantity presents a significant obstacle to engineering students' algebraic reasoning. Furthermore, it appears that this obstacle is increased by a one semester physics course. Results on item VIII are consistent with the findings of Cohen and Kanim,[45] and performance shows no improvement after one semester of mechanics. As a group, the results on items VIIc, VIId, and VIII suggest that even engineering students struggle to symbolize in simple physical contexts.

15Comparison of the fraction and variable versions of item VII (see Fig. 3) reveals that substituting even just one generalized variable, represented by a letter, for a numeric value inhibits the appropriate reasoning at least as much as the replacement of whole numbers with decimals numbers. Our results here are consistent with those of Torigoe and Gladding [47] in which they compared algebraic computations with numeric ones. Our work, however, extends the effort to disentangle the source of students' difficulty. While most instructors would probably agree that a purely symbolic equation is more challenging for students than one that has just one symbol, it is perhaps surprising that only a single generalized variable can hinder student reasoning to the extent we have measured here. For nearly half of the population tested, the presence of a single variable quantity disrupts reasoning the students were successful with in the context of whole number quantities.

Also consistent with published results,[45,46] we find that students in this population struggle to interpret symbols appropriately when asked to generate a simple mathematical statement from a verbal one. On Item VIII, fewer than one-half of the students selected the equation that matches the given verbal statement. The most commonly selected incorrect choice involved a reversal of the variables. We speculate that it is not that students cannot do algebra, or even manipulate algebraic statements necessarily, but that even after a semester of calculus-based physics they are already uneasy at the problem statement – there are many students who are destabilized once they see the first variable quantity.

D. Summary discussion

Our study reveals that many students don't think about quantities in physics as the cognitive blended objects that they are. During the interviews, students tended to focus on the numeric values as the mathematical objects to be operated on, and treated the units as add-ons. In contrast, physicists reflexively join the number and its unit – seeing quantity as a cognitive blend of numbers and units. The units are a foundational part of sense making that often guide mathematical thinking. Our results suggest that many students emerge from an introductory mechanics course lacking facility with this type of reasoning.

The students' unblended approach to quantification renders other cognitive work more difficult. The merits of a dimensionless comparison may not be clear without first considering units as inherent parts of quantities (i.e., rather than "the thing that costs you points for leaving out"). Students struggle to generate ratio as an appropriate comparison; their tendency to adopt a difference strategy on the squareness and force vectors items, and, when interviewed, their struggle to understand why this is not a generally useful approach, indicate that they have not internalized the notion of quantity in the way we would hope after having taken a physics course. For many students, the cognitive domains of mathematical procedure and physical sense making are distinct and isolated, rather than blended.

A ratio involves the comparison of two quantities. The associated reasoning, when nascent, can be inhibited if those quantities are numerically complex or abstract. The effects of numerical complexity are especially notable when quantities are represented by general variables. We see that mathematically strong students struggle to conceptualize and generate a simple proportional relationship between two variable quantities (number of heffalumps and number of woozles) in spite of having completed a course in which much of what they have written in their lecture notes consists of symbolic representations of proportional relationships. It appears that many have functioned as good scribes, but only a portion have internalized the fundamental reasoning.

A weighted average of the assessment items yields a correct post test response rate of just over 50%, suggesting that a typical introductory calculus-based mechanics course has only limited ability to help even well prepared students learn to reason consistently about ratio and proportion. This finding is disturbing, given that proportional reasoning is fundamental to all of physics, and that instructors commonly model it when they teach.

From the cognitive blending perspective, algebraic reasoning is inextricably bound to the context in which it is being used. We argue that this cognitive space is entirely homogeneous, in the sense that in a physics course, "doing algebra" is not separate from "doing physics." It is not uncommon for experts to make errors similar to those of students when they initially answer items I, II, VII (all versions) and VIII. Experts, however, expect an answer to make sense in context, and employ a variety of tactics to evaluate answers before they consider the problem to be completed. It seems likely that this tight binding of algebraic reasoning and physical sense making is not part of what all students are learning from experts when they take a physics course.

## VI. Recommendations for instruction

We conclude that successfully completing the mathematics prerequisites for a physics course primarily prepares students for the *procedural* aspects of the mathematics necessary for following physics explanations, but not for the mathematical reasoning needed for physical sense making. Physicists, because of their deep knowledge of the contextualized use of mathematics within the discipline, are best positioned to help students develop this reasoning. Yet instructors may be unintentionally creating barriers to success for many students in introductory physics by using "pre-fabricated" explanations when modeling mathematical sense making in physics, without attending to the fabrication process explicitly.

There has not yet been sufficient research to establish valid models for improving instruction. For students, building and coordinating cognitive resources in physics involves simultaneously developing a sense of what new and abstract quantities actually represent, and learning a new process of reasoning algebraically with abstract dimensioned quantities. Much instructional effort and research has gone into the service of the former; we emphasize the need for additional research on the latter. Research-validated methods and materials for facilitating learning the fabrication process are needed.

We offer the following recommendations for instruction based on the assumption that explicitly acknowledging and addressing the cognitive challenges of the way we quantify in physics can result in productive learning. We suggest that it is essential for students to participate in the generation of simple mathematical ideas in order to learn physics thinking. Instructors can facilitate learning to use algebra with poorly understood quantities through explicit cognitive apprenticeship - not just about teaching "how to", but facilitating discovering why. We provide some specific suggestions below that reflect the experience of the authors over the course of this work and that are driven by the student difficulties revealed and supported by this work.

*Quantification, general structural reasoning and context.* Our results reveal that most engineering students do not reliably reason with the units of physical quantities - even at the end of an introductory physics course. In addition, they often fail to generalize basic mathematical structures in physics. Engaging students in the process of deciding which of the four arithmetic procedures makes sense while introducing new quantities and laws might help bridge students' understanding of why calculus is not only necessary, but also helpful. For example, understanding generally that conservation laws involve summing up quantities (which are themselves products of other quantities) can help students decode the symbolizing that goes into the integral version of the work-energy and impulse-momentum theorems.

The use of a ratio for comparison is of particular interest in physics contexts. We suggest that explicit cognitive apprenticeship can enhance student decision-making associated with comparisons. Instruction can be modified to "open up" the decision-making surrounding algebraic reasoning and assumptions, with a variety of (dimensioned) physical quantities, through targeted clicker questions in lecture and modifications to collaborative learning curricular materials.

*Symbolizing and numerical complexity.* We suggest that instructors recognize the cognitive burden that quantity can have as they lecture and construct exams, and that abstract quantities get mastered first using whole, single-digit numbers. Using whole numbers in physics, where the quantities represent real measured values, presents a particular problem in many contexts (e.g., E&M, gravitation) and we suggest in those contexts to refer to invented "charge units", "mass units" and "distance units" as the students are learning to master the mathematical structure of the reasoning. We suggest that use of general variables should occur only after practice with simple numbers.

Further research and curriculum development is needed to better understand how to help students develop generalized structural reasoning, facility with physics quantities and how to symbolize these quantities. We view recent increased interest in physics students' mathematical reasoning and increased dialogue with mathematics education researchers as important trends in physics education research. We believe these trends will result in improved understanding of how experts use mathematics for sense making in physics and how student thinking aligns with and diverges from expert mathematization. The development of instructional approaches that can help bridge the gaps will naturally follow. The work described in this paper is intended to contribute to these goals.

**Acknowledgments**

This work has been supported in part by the National Science Foundation, under DUE #1045250, #1045227, and #1045231. The authors are grateful for discussions with Patrick Thompson and Dan Schwartz, whose work has deeply informed our understanding of conceptualization of mathematics. We thank Eugenia Etkina for valuable feedback on the manuscript in its early stages, and Eleanor Sayre and two anonymous reviewers for suggestions that helped clarify the essence of this work. We thank Eugene Geis for assistance with data analysis.

1. Hans Freudenthal, *Mathematics as an educational task*, (D. Reidel, Dordrecht, 1973).

2. Adrian Treffers, *Three dimensions: A model of goal and theory description in mathematics instruction-The Wiskobas Project*, H. Vonk *et al.* translators, (D. Reidel Dordrecht, 1987).

3. Edward F. Redish and Eric Kuo, "Language of Physics, Language of Math: Disciplinary Culture and Dynamic Epistemology," Science & Education **24** (5/6), 561-590 (2015).

4. Olaf Uhden *et al.*, "Modelling mathematical reasoning in physics education," Science & Education **21** (4), 485-506 (2012).

5. Gilles Fauconnier and Mark Turner, *The way we think: Conceptual blending and the mind's hidden complexities*, (Basic Books, 2008).

6. Thomas J. Bing and Edward F. Redish, "The cognitive blending of mathematics and physics knowledge," in *2006 Physics Education Research Conference*, edited by L. McCullough, L. Hsu, and P. Heron (AIP, New York, 2007), 26-29.

7. Michael C. Wittmann and Katrina E. Black, "Mathematical actions as procedural resources: An example from the separation of variables," Physical Review Special Topics-Physics Education **11** (2), 020114 (2015).

8. Jonathan Tuminaro, "A cognitive framework for analyzing and describing introductory students' use and understanding of mathematics in physics," Doctor of Philosophy, University of Maryland, (2004).

9. Michael M. Hull, Eric Kuo, Ayush Gupta, and Andrew Elby, "Problem-solving rubrics revisited: Attending to the blending of informal conceptual and formal mathematical reasoning," Physical Review Special Topics-Physics Education Research **9** (1), 010105 (2013).

10. Jennifer A. Czocher, "Toward an understanding of how engineering students think mathematically," Education; Doctor of Philosophy, the Ohio State University, (2013).

11. Steven R. Jones, "Areas, anti-derivatives, and adding up pieces: Definite integrals in pure mathematics and applied science contexts," The Journal of Mathematical Behavior **38**, 9-28 (2015).

12. Steven R. Jones, "Understanding the integral: Students' symbolic forms," Journal of Mathematical Behavior **32** (2), 122-141 (2013).

13. Ricardo Karam, "Framing the structural role of mathematics in physics lectures: A case study on electromagnetism," Phys.Rev.ST Phys.Educ.Res. **10**, 010119 (2014).

14. Eric Kuo, E. *et al.*, "How students blend conceptual and formal mathematical reasoning in solving physics problems," Science Education **97** (1), 32-57 (2013).

15. D. C. Meredith and K. A. Marrongelle, "How students use mathematical resources in an electrostatics context," American Journal of Physics **76** (6), 570-578 (2008).

16. N. Sanjay Rebello *et al.*, "Transfer of learning in problem solving in the context of mathematics and physics," in *Learning to solve complex scientific problems*, edited by David H. Jonassen (Routledge, 2007) 223-246.

17. Bruce L. Sherin, "How students understand physics equations," Cognition and instruction **19** (4), 479-541 (2001).

18. Joshua Von Korff and N. Sanjay Rebello, "Distinguishing between "change" and "amount" infinitesimals in first-semester calculus-based physics," American Journal of Physics **82** (7), 695-705 (2014).

19. Joshua Von Korff *et al.*, "Student Epistemology About Mathematical Integration In A Physics Context: A Case Study," in *2013 Physics Education Research Conference*, edited by Paula V. Engelhardt, Alice Churukian, and Dyan L. Jone*s*, (AIP 2013) 353-356.

20. Marcos D. Caballero *et al.*, "Unpacking students' use of mathematics in upper-division physics: Where do we go from here?" European Journal of Physics **36** (6), 065004 (2015).

21. Alan H. Schoenfeld, *Mathematical problem solving*, (Academic Press, New York, 1985).

22. David Hammer *et al.*, "Resources, framing, and transfer," in *Transfer of learning from a modern multidisciplinary perspective* edited by Jose P. Mestre (IAP, Greenwich, 2005), 89-120.

23. David Hammer, "Student resources for learning introductory physics," American Journal of Physics **68**, S52-S59 (2000).

24. Michael C. Wittmann, "Understanding coordinated sets of resources: An example from quantum tunneling," in *Proceedings of the Fermi School in Physics Education Research*, Edited by Edward F. Redish and Matilde Vicentini, (IOS Press, Varenna, Italy, 2004).

25. Arnold. B. Arons, *A guide to introductory physics teaching,* (Wiley, 1990).

26. Arnold B. Arons, "Cultivating the capacity for formal reasoning: Objectives and procedures in an introductory physical science course," American Journal of Physics **44** (9), 834-838 (1976).

27. Robert Karplus, Steven Pulos and Elizabeth K. Stage, "Early adolescents' proportional reasoning on 'rate'problems," Educational studies in Mathematics **14** (3), 219-233 (1983).

28. Robert Karplus and R. W. Peterson, "Intellectual Development Beyond Elementary School II*: Ratio, A Survey," School Science and Mathematics **70** (9), 813-820 (1970).

29. Eddie M. Gray and David O. Tall, "Duality, ambiguity, and flexibility: A" proceptual" view of simple arithmetic," Journal for Research in Mathematics Education , **25**, 116-140 (1994).

30. Alan H. Schoenfeld, "Learning to think mathematically: Problem solving, metacognition, and sense making in mathematics," in *Handbook of research on mathematics teaching and learning* , edited by Douglas A Grouws (MacMillan, 1992) 334-370.

31. Patrick W. Thompson and Luis A. Saldanha, "Fractions and multiplicative reasoning," in *Research companion to the principles and standards for school mathematics* , edited by Jeremy Kilpatrick, (NCTM, 2003) 95-113.

32. Francoise Tourniaire and Steven Pulos, "Proportional reasoning: A review of the literature," Educational studies in mathematics 16 (2), 181-204 (1985).

33. Patrick W. Thompson *et al*., Schemes for thinking with magnitudes: A hypothesis about foundational reasoning abilities in algebra, in *Epistemic algebra students: Emerging models of students' algebraic knowing* , edited by K. C. Moore,L.P.Steffe & L.L.Hatfield, (University of Wyoming, Laramie, WY, 2014) 1-24.

34. Andrew Boudreaux, Stephen Kanim and Suzanne Brahmia, "Student facility with ratio and proportion: Mapping the reasoning space in introductory physics," J arXiv preprint arXiv:1511.08960 , (2015).

35. Martin A. Simon and Glendon W. Blume, "Mathematical modeling as a component of understanding ratio-as-measure: A study of prospective elementary teachers," The Journal of Mathematical Behavior **13** (2), 183-197 (1994).

36. Patrick W. Thompson, Quantitative reasoning and mathematical modeling, in *New perspectives and directions for collaborative research in mathematics education*, edited by L. L. Hatfield,S.Chamberlain & S.Belbase, (University of Wyoming, Laramie, WY, 2011) 33-57.

37. Hans-Christoph Nuerk *et al*., "Extending the Mental Number Line," Zeitschrift für Psychologie **219** (1), 3-22 (2011).

38. Catherine Thevenot and Jane Oakhill, "The strategic use of alternative representations in arithmetic word problem solving," The Quarterly Journal of Experimental Psychology Section A **58** (7), 1311-1323 (2005).

39. For a comprehensive overview, see *Symbolizing and communicating in mathematics classrooms: Perspectives on discourse, tools, and instructional design*, edited by Paul Cobb, Erna Yackel and Kay McClain, (Routledge, New York, 2000).

40. María Trigueros and Sonia Ursini, "Structure Sense and the use of variable," in the *Proceedings of the Joint Meeting of PME 32 and PME-NA XXX. Vol. 1. México: Cinvestav-*

*UMSNH,* edited by Olimpia Figueras José Luis Cortina Silvia Alatorre Teresa Rojano Armando Sepúlveda 4:337-344 (Morelia, Mexico, 2008).

41. María Trigueros and Sonia Ursini, "First-year undergraduates' difficulties in working with different uses of variable," CBMS issues in mathematics education **8**, 1-26 (2003).

42. Randolph A. Philipp, "The Many Uses of Algebraic Variables," The Mathematics Teacher **85** (7), 557-561 (1992).

43. Edward F. Redish, "Problem solving and the use of math in physics courses," arXiv preprint physics/0608268 , (2006).

44. Carolyn Kieran, "Concepts associated with the equality symbol," Educational studies in mathematics **12** (3), 317-326 (1981).

45. Elaine Cohen and Stephen E. Kanim, "Factors influencing the algebra "reversal error"," American journal of physics **73**, 1072 (2005).

46. John Clement, Jack Lochhead and George S. Monk, "Translation difficulties in learning mathematics," American mathematical monthly **88**, 286-290 (1981).

47. Eugene Torigoe, "How numbers help students solve physics problems," J arXiv preprint arXiv:1112.3229, (2011).

48. Patricia M. Heller *et al*., "Proportional reasoning: The effect of two context variables, rate type, and problem setting," Journal of Research in Science Teaching **26** (3), 205-220 (1989).

49. David Hestenes, Malcolm Wells and Gregg Swackhamer, "Force concept inventory," The Physics Teacher **30** (3), 141-158 (1992).

50. Douglas R. Mulford and William R. Robinson, "An inventory for alternate conceptions among first-semester general chemistry students," J.Chem.Educ. **79** (6), 739 (2002).

51. Laura McCullough, "Gender, context, and physics assessment," Journal of International Women's Studies **5** (4), 20-30 (2004).

52. John Stewart, Heather Griffin and Gay Stewart, "Context sensitivity in the force concept inventory," Phys.Rev.ST Phys.Educ.Res. **3,** 010102 (2007).